# On Object-Orientation

*Bob Diertens*

section Theory of Computer Science, Faculty of Science, University of Amsterdam

*ABSTRACT*

Although object-orientation has been around for several decades, its key concept abstraction has not been exploited for proper application of object-orientation in other phases of software development than the implementation phase. We mention some issues that lead to a lot of confusion and obscurity with object-orientation and its application in software development. We describe object-orientation as abstract as possible such that it can be applied to all phases of software development.

## 1. Introduction

The use of the concepts of object-orientation (OO) in programming predates the existance of programming languages. In the 1960s the programming language Simula appears, which is considered the first language supporting OO and developed into the language Simula 67 [7]. This language was used as a platform for the development of the programming language Smalltalk [8] in the 1970s. In the 1980s C++ [13] was introduced, bringing the concepts of Simula into the C programming language. From the 1990s object oriented programming became a dominant style for implementing complex programs consisting of interacting components.

Along side of object oriented programming (OOP), object orientation has been applied in design (OOD) of software systems, and in analysis (OOA) of the requirements for a software system. Over the years, methodologies for software construction have become more and more structured. These methodologies were either data-oriented, function-oriented, or both although still seperated. With the rise of OO in programming the need for OO in design, and later OO in analysis, came into existence. A software system described in terms of functions and/or data had to be mapped onto a description in terms of objects. Using OO in earlier phases can smooth transitions from one phase to another.

A brief history of the object-oriented approach to software development is given in [5], together with a survey of object-oriented methodologies. It mentions among others the work of Shlaer and Mellor [12], Coad and Yourdon [6], Jackson [9], Booch [1] [2] [3], Jacobson [10], Wirfs-Brock et al. [14], and Rumbaugh et al. [11].

Although a lot of work has been performed in the field of OO, there is still a lot of confusion and obscurity in this world. This is partly due to the use of different terminology and the use of different semantics for the same terms. Another aspect is that the fundamentals of OO are often explained in terms of features provided by OOP languages. With as result that design on higher levels of abstraction is expressed on the level of implementation.

With OO we can abstract from the implementation of objects, enabling us to concentrate on the behaviour of objects and their relations with each other. With abstraction we can deal with the complexity of the system. But this aspect of OO is seldom used. Instead, old methods are used and packaged in objects, thereby increasing the complexity of the system.

In this paper we describe OO the way we see it and that can be applied to all phases of software development. In order to apply OO in other phases of software development than implementation we have to define OO without using features offered by OOP languages. We have to have an abstract model of OO that can be applied to all these phases.

In the next section we briefly describe some issues with OO as currently used in software development methodologies. We set out our thoughts on OO in section 3 and apply this to software development in section 4.

## 2. Issues with Object-Orientation

We describe some issues concerning OO that lead to a lot of confusion and obscurity in this field. This list is by no means complete, but is intended to pinpoint some of the problems due to the way OO is currently applied by a lot of practitioners.

**Features**
When people are asked for the fundamentals of OO, they often reply with a list of features provided by OOP languages instead of what OO is truly about. The features mentioned mostly are classification, inheritance, polymorphism, encapsulation, and abstraction. For OO these features are irrelevant, apart from abstraction but the term is wrongly used here. To describe OO in terms of features provided by OOP languages that support OO leads to the conclusion that for a programming language to be OO, it has to support these features. This circular reasoning is certainly not helpful for a good understanding of what OO is truely about.

Emphasis on the features of OOP languages, such as classification, inheritance, and polymorphism, leads to specification of data objects and distracts from the OO concepts of behaviour abstraction. Also on higher levels of abstraction in design there is less need for these features, with as result that the design is directly done on the level of programming. It seems that the use of features provided by OOP languages has become the goal.

**Abstraction and Generalization**
There is a lot of confusion over abstraction and generalization, or rather, they are interchanged. But abstraction and generalization are definitely not the same. With abstraction some detail is left out that is considered not important in a description on an higher level of abstraction. With generalization that detail is not left out, but described in a general way on the same level of abstraction.

Consider the following example where we have a red, green, and blue object. We can describe these objects in general in terms of an object with a particular color. With abstraction we describe these objects as an object without the mentioning of a color at all.

**Encapsulation and Information Hiding**
Encapsulation and information hiding are often interchanged or used with the same meaning. Encapsulation of an object prevents communication with that object in other ways than the defined ones and does not hide how an object does things. Information hiding makes it impossible to see how an object does things, but does not prevent communication with that object in certain ways.

**Description**
Objects are often described in terms of data and a set of functions. The Unified Modeling Language (UML) [4] is advocated as a language for this in all phases of software development. With the description of objects in terms of data and functions, UML hardly rises from the level of OOP languages. Although the details of data and functions are left out, there is no abstraction from the implementation of objects. This way of describing objects are thus useless on higher levels of abstraction in software design. Objects should be described by their behaviour, instead of in terms of data and functions.

## 3. Object-Orientation

OO is a modelling paradigm for describing objects and their relationships. Objects and relations are supposed to stand close to real world concepts. The real world is the world we are implementing, that is a level of abstraction in the design or a requirements specification. The real world is a future world in which the system under development takes part.

The real world is also an abstract world. It is of no concern how something works, only what it does. This abstraction is key in OO. However, OO is often easily replaced with OOP. But an implementation in an OOP language is no more than an example of this modelling at the lowest level of abstraction of the design.

Because of this replacement, OO is explained by describing what an particular OOP language has to offer. To define a model of OO that can be applied in several phases of software development we have to define this with as much abstraction as possible. We give a description of the fundamentals of OO, techniques to support the fundamentals, and features based on the techniques. Note that only the fundamentals are necessary for object-oriented modelling, some support can be nice, and features are mostly only used on the lowest levels of abstraction.

## *3.1 Fundamentals*

OO can be seen as a kind of technique of organizing a system in terms of objects and their relations. It is supposed to stand closer to the *real* world as opposed to techniques predating OO. Its characterist is the distinction between the observable behaviour of objects and the implementation of the behaviours.

**Objects**
An object has the following characteristics.

> **state**
> for recording the history of an object upon which future behaviour can be based.
>
> **behaviour**
> the observable effects based on its state and the relations with other objects.
>
> **identity**
> as known by other objects, either by name or by reference.

**Relations**
Relations between objects are expressed by interactions in the form of message passing.

**Abstraction**
Manipulation of an object can only be done through its relations with other objects. Thereby hiding the implementation of its behaviour and the recording of its state. It is only important what an object does, not how an object does it.

## *3.2 Support*

An object-oriented language for modelling systems on a particular level of abstraction has to support the fundumentals of OO and possibly even enforce these fundamentals. Support can be provided in the following forms.

**Types**
An object type is a container in which the state and the behaviour(s) for an object are defined.

**Message Passing**
The way messages are passed between objects can be supported in more than one form.

**Encapsulation**
Encapsulation prevents objects from relating to each other in other ways then the provided forms of message passing.

**Information Hiding**
Hiding of information about an object can be done by deliberately making this information inaccessible.

## *3.3 Structures*

Based on the techniques supporting OO structures can be formed. Such structures behave as objects themselves, characterizing the concepts of OO.

**Type Composition**
The basic idea of composition is to build complex object types out of simpler ones. Besides that objects can be built up from ways to define state and behaviour as provided by the modelling language, objects can also be built up from other object types. The latter can be done in the following forms.

**reference**
    An object type can reference an object of a particular object type.

**inclusion**
    An object type can include another object type.

To obey the OO fundamentals of keeping behaviour and implementation of an object distinct, a modelling language has to hide the composition of an object type. This can be achieved by making the elements of the object type acquired through composition available either only from within the object type, or from outside the object type but as it were elements of the object type itself.

Objects composed in this way are vertical related with the objects they are composed of.

**Object Composition**
Several inter-related objects form a cluster that when abstracted from the inter-relations acts as a single object. Objects that take part in this composition are horizontal related with eachother.

**Abstract Object Types**
An abstract object type is an object type described in terms of objects representing elements of the abstract object type for which the type(s) have to be filled in on a lower level of abstraction. An abstract object type can also be turned into a generic object type on a lower level of abstraction, with parameters for the object types representing the elements.

## 4. Application of OO in Software Development

It is often said that OO stands closer to the real world as opposed to the techniques predating OO. The real world is actually an abstract world that stands far from OO applied in a software system. If we want to properly apply OO in the implementation of a software system we have to close the gap between real world objects and objects in the implementation. Therefor it is logical to apply OO in the earlier phases of software development.

Software development in the past moved from ad hoc methods to more structured methods. The structured methods can be characterised as either data-oriented or function-oriented, but were usually a mixture of both. The object-oriented paradigm abstracts from how something is achieved by an object to what it achieves, and emphasises the interaction between objects. How something is achieved by an object can thus be deferred until later in the development process. Unfortunately, this characteristic is less used. Instead, a data oriented class hierarchy is developed in an early phase of development, even in the analysis phase.

In the following we describe some issues with the current application of OO in several phases of software development and how OO our view can be used in these phases.

**Analysis**
OOA often results in designing the system. This is largely due to trying to define a class hierarchy for data. With this, objects are represented as data and so the fundamentals of OO are not obeyed.

Actually, in analysis OO is of no use for describing the software system. Should OO be used for describing what the software system has to do, we are actually designing the software system. However, OO can be used for describing the interaction of the software system with the environment in which it will be deployed. The software system itself is an object in this environment, or actually in this future environment.

**Design**
In most cases, the design concentrates around the development of class hierarchies. Moreover, these class hierarchies are expressed directly in an OOP language or in a language on a too low level of abstraction. Instead it should focus on abstraction in order to deal with the complexity of the system. This can be achieved by developing abstract models of the system, each a refinement of the other. OO is extremely usefull in this, since it enables abstraction from internal behaviour and implementation. This also makes the study of different refinements for a particular object possible.

**Analysis & Design**

Analysis and design can be overlapping phases because some design decision reveal a need for further analysis or analysis depend on how certain parts of the system can be implemented. But most of all, design experiments can give information on incomplete or vague requirements.

OO supports this overlapping of phases because abstraction makes it possible to delay design decisions concerning the internal working of objects. Abstraction also allows for experimenting on different levels of abstraction of design.

**Programming**

An object-oriented design is easier to map onto an object-oriented implementation than a design that is not object-oriented. An object-oriented design also increases the reusability of code.

There is no particular programming language needed to obey the fundamentals of OO when programming. However, it can be convenient to program in an object-oriented style in a language that supports OO. Currently, in the use of OOP languages there is a high stress on developing class hierarchies to make reuse of code possible. It is not the use of classes that leads to reuse, but abstraction and thus a thorough design that leads to reuse.

**Maintenance**

Application of OO in all phases of software development makes maintenance of the software easier. It is easier to see at which level of abstraction certain design decisions have been made. Also, it is easier to see at which level changes have to be incorporated into the design and which parts of the software are affected by these changes.

## 5. Conclusions

After studying the available literature on OO and its application in software development we concluded that there is a lot of confusion and obscurity in this world. We have described some issues with OO leading to this. We also concluded that abstraction, the key concept of OO, is seldom used in dealing with the complexity of software systems.

In order to apply OO in all phases of software development we have described the fundamentals of OO and ways to support these fundamentals, with as much abstraction as possible. Furthermore, we described how to apply our view on OO in software development.

We hope that it contributes to a thorough understanding of OO and how it must be applied in software development. We intend to use it as a base in the development of support for modelling software systems at different level of abstraction, including implementation models preserving the parallelism objects have by nature.